\documentclass[12pt,a4paper]{article}
\usepackage{a4wide}
\usepackage{latexsym}
\usepackage{epsf}
\usepackage{amssymb}

\makeatletter
\@addtoreset{equation}{section}
\makeatother

\begin{document}

\begin{flushright}
\small
IFT-UAM/CSIC-00-38\\
KUL-TF-2000/28\\
{\bf hep-th/0012032}\\
December $5$th, $2000$
\normalsize
\end{flushright}

\begin{center}

%title

\vspace{.7cm}

{\LARGE {\bf An $SL(3,\mathbb{Z})$ Multiplet of 8-Dimensional}}\\
\vspace{.3cm}
{\LARGE {\bf Type~II Supergravity Theories and the}}\\
\vspace{.3cm}
{\LARGE {\bf Gauged Supergravity Inside}}

\vspace{1.2cm}

%authors
{\bf\large Natxo Alonso-Alberca},${}^{\spadesuit}$
\footnote{E-mail: {\tt Natxo.Alonso@uam.es}}
{\bf\large Patrick Meessen}${}^{\diamondsuit}$
\footnote{E-mail: {\tt Patrick.Meessen@fys.kuleuven.ac.be}}
\vskip 0.3truecm
{\bf\large and Tom\'as Ort\'{\i}n}${}^{\spadesuit\clubsuit}$
\footnote{E-mail: {\tt tomas@leonidas.imaff.csic.es}}
\vskip 1truecm

${}^{\spadesuit}$\ {\it Instituto de F\'{\i}sica Te\'orica, C-XVI,
Universidad Aut\'onoma de Madrid \\
E-28049-Madrid, Spain}

\vskip 0.2cm 
${}^{\diamondsuit}$\ {\it Instituut voor Theoretische
Fysica, Katholieke Universiteit Leuven, Celestijnenlaan~200D, B-3001
Leuven, Belgium.}

\vskip 0.2cm
${}^{\clubsuit}$\ {\it I.M.A.F.F., C.S.I.C., Calle de Serrano 113 bis\\ 
E-28006-Madrid, Spain}

\vspace{.7cm}

%%%%%%%%%%%%%%%%%%%%%%%%%%%%%%%%%%%%%%%%%%%%%%%%%%%%%%%%%%%%%%%%%%%%%%

{\bf Abstract}

\end{center}

\begin{quotation}

\small

The so-called ``massive 11-dimensional supergravity'' theory gives,
for one Killing vector, Romans' massive 10-dimensional supergravity in
10 dimensions, for two Killing vectors an $Sl(2,\mathbb{Z})$ multiplet
of massive 9-dimensional supergravity theories that can be obtained by
standard generalized dimensional reduction type~IIB supergravity
and has been shown to contain a gauged supergravity.

We consider a straightforward generalization of this theory to three
Killing vectors and a $3\times 3$ symmetric mass matrix and show that
it gives an $Sl(3,\mathbb{Z})$ multiplet of 8-dimensional supergravity
theories that contain an $SO(3)$ gauged supergravity which is, in some
way, the dual to the one found by Salam and Sezgin by standard
generalized dimensional reduction.

\end{quotation}

\newpage

\pagestyle{plain}

%%%%%%%%%%%%%%%%%%%%%%%%%%%%%%%%%%%%%%%%%%%%%%%%%%%%%%%%%%%%%%%%%%%%%%
\section*{Introduction}
%%%%%%%%%%%%%%%%%%%%%%%%%%%%%%%%%%%%%%%%%%%%%%%%%%%%%%%%%%%%%%%%%%%%%%

Massive and gauged supergravity theories are fascinating theories in
which there has been much interest recently due to their connections,
as effective string theories, with Maldacena's AdS/CFT correspondence
conjecture.

It is clear that there is a lack of understanding of many basic
features of these theories. One of the main puzzles is their origin
from compactification of higher-dimensional theories.  This is the
case of Romans' massive $N=2A,d=10$ supergravity \cite{kn:Ro2} which,
according to the standard lore, should have its origin in
11-dimensional supergravity but so far it has been impossible to
obtain from it by standard methods.

To explain the 11-dimensional origin of Romans' massive $N=2A,d=10$
supergravity a massive 11-dimensional theory, containing a mass
parameter and an explicit Killing vector in its action, was proposed
in Ref.~\cite{kn:BLO}.  Dimensional reduction of the theory in the
direction of the Killing vector gives Romans' theory.  The
construction of this theory was based on the fact that one could
derive the effective action of some supersymmetric solitons of Romans'
theory from 11-dimensional gauged sigma models, in which the gauging
was associated to a single Killing isometry of the background
\cite{kn:O} and the theory was interpreted as the theory resulting
from 11-dimensional supergravity in a background consisting of a {\it
  KK-brane}, with 9 spatial worldvolume dimensions and one special
isometric direction that had to be compact, just as the special
isometric direction of the KK monopole has to be compact.\footnote{The
  effective theory of the KK monopole is also a gauged sigma model of
  the same kind \cite{kn:BJO2}.}  We will have more to say about the
interpretation of this kind of theories later.

In Ref.~\cite{kn:MO} the massive 11-dimensional theory of
Ref.~\cite{kn:BLO} was generalized in order to maintain some sort of
S~duality covariance of theory compactified on $T^{2}$: the origin of
the S~duality ($Sl(2,\mathbb{Z})$) symmetry of the 9-dimensional
theory is simply the reparametrization-invariance of the internal
manifold. This symmetry was broken by the presence of the KK-brane in
one of the compact directions but one could introduce a second
KK-brane in the orthogonal internal direction. Then,
$Sl(2,\mathbb{Z})$ transformations bring us from one background with
two given KK-branes to an S~dual background.  The generalized
11-dimensional massive theory has 2 commuting Killing vectors
$\hat{\hat{k}}_{(m)}{}^{\hat{\hat{\mu}}}$ $n=1,2$ appearing explicitly
in the action and a symmetric mass matrix $Q^{mn}$ whose entries could
be related to brane charges. Dimensional reduction of this theory in
the direction of the two Killing vectors gives the 9-dimensional
massive theories of Ref.~\cite{kn:MO} that can also be obtained by GDR
from type~IIB supergravity.

Recently, Cowdall \cite{kn:cow} realized that this 9-dimensional
contains the long sought-after $D=9$ $N=2$ gauged supergravity theory.
To show this explicitly, he had to perform a few field redefinitions
and eliminate 2 of the 3 vector fields through massive gauge
transformations ({\em i.e.}~letting them be eaten up by the 2 2-forms
that then become massive) and then it could be seen that the resulting
action has a $SO(2)$ gauge symmetry.  This $SO(2)$ is also the
R-symmetry of the $d=9$ $N=2$ supertranslation algebra, so that the
identification of the found theory with the gauged supergravity theory
is obvious.  It should be stressed that the local $SO(2)$ symmetry
does not arise in any special singular limit but that it is already
present in the original massive theory albeit in a rather
unconventional form due to the presence of additional massive
parameters.

The question arises immediately as to whether this new connection
between massive and gauged supergravities is more general and why
there is such relation at all. Our goal in this work is to investigate
this relationship by exploring new examples of massive and gauged
supergravities. At the same time we will study their 11-dimensional
origin.

Let us first describe the new massive supergravity theories we present
in this paper and how we are going to obtain them. It should be clear
that the massive 11-dimensional theory of Refs.~\cite{kn:BLO,kn:MO}
can be further generalized by allowing the index $n$ to go from $1$ to
$3$, for instance, {\em i.e.}~having $3$ commuting Killing vectors in
$11$ dimensions (all the fields being independent of the $3$
associated coordinates) and a $3\times 3$ symmetric mass matrix
$Q^{mn}$.  This theory would give a standard fully general-covariant
theory in $d=8$ upon reduction in the directions of the $3$ Killing
vectors that now appear explicitly in the action. This would be a
massive theory (in fact a full $Sl(3,\mathbb{Z})$ multiplet of massive
theories,\footnote{The classical theory has continuous duality groups:
  $Sl(3,\mathbb{R})\times Sl(2,\mathbb{R})$ that are broken by
  quantum-mechanical effects to the discrete subgroups
  $Sl(3,\mathbb{Z})\times Sl(2,\mathbb{Z})$. To the same conclusion
  one arrives by considering the preservation of the periodic boundary
  conditions of the coordinates in the internal dimensions. All our
  results will be classical and we will not pay much attention to
  these details unless strictly necessary.} with the mass matrix
$Q^{mn}$ transforming under $Sl(3,\mathbb{Z})$), but we may ask
ourselves whether such a theory really is a supergravity theory, given
that only the bosonic sector was considered in
Refs.~\cite{kn:BLO,kn:MO}. It should be clear that the same theory
could be obtained by standard procedures known to preserve
supersymmetry: first, we can perform the standard dimensional
reduction of the massive 9-dimensional supergravity theories of
Ref.~\cite{kn:MO} to 8 dimensions. The resulting theory is a massive
supergravity theory with a $2\times 2$ mass matrix. The massless
theory has an $Sl(3,\mathbb{Z})\times Sl(2,\mathbb{Z})$ duality group
and we could simply perform general $Sl(3,\mathbb{Z})$ transformations
in the massive theory to introduce new mass parameters.\footnote{This
  is analogous to the procedure used to construct new
  duality-covariant families of black-hole solutions. see {\em
    e.g.}~\cite{kn:KO}.} This procedure should preserve supersymmetry
and give the same massive 8-dimensional theory, but it is difficult to
implement in practice. Our proof that the theory we obtain is indeed a
good standard supergravity theory will be to establish its relation to
a well-known supergravity theory as we are going to explain shortly.

Does this new massive supergravity theory contain a gauged
supergravity?  In $d=8$ the R-symmetry group of the superalgebra is
$SO(3)$ which is just the local invariance of the scalars in
$\mathcal{M}\in Sl(3,\mathbb{R})/SO(3)$. An $SU(2)$ $d=8$ gauged
supergravity was obtained by Salam and Sezgin by means of
Scherk-Schwarz generalized dimensional reduction \cite{kn:SS} from
11-dimensional supergravity in Ref.~\cite{kn:SS}. Then, a gauge theory
exists and we want to know if our massive 8-dimensional theory
contains it. Since both 8-dimensional theories have very different
11-dimensional origins this is a non-trivial test of our ideas.

We are going to show that indeed the massive theory contains a gauge
theory but we are going to go further: the massive theory is really
nothing but an $SO(3)$ gauged theory in disguise in which a
non-standard basis of the $so(3)$ Lie algebra is used and in which
St\"uckelberg fields are present so the mass terms of some fields look
like field strengths. The theory is directly $SO(3)$ gauge invariant
and, at the same time it has massive gauge invariance. There is no
need to take any particular limit and removing the St\"uckelberg
fields and going to a standard basis of $so(3)$ is just a matter of
esthetics and choice and, at the same time, of maintaining
$Sl(3,\mathbb{R})$ covariance in the sense of
Refs.~\cite{kn:MO,kn:KM}.

What is the relation of this theory to Salam and Sezgin's? The theory
turns out to be very similar but, there are two subtle, but important,
differences: first of all, the $SO(3)$ vectors in Salam and Sezgin's
(SS) theory are the Kaluza-Klein (KK) vectors while in the theory that
we are going to derive from the 11-dimensional construction of
Refs.~\cite{kn:BLO,kn:MO}, the $SO(3)$ vectors are associated to
membranes wrapped on 2 cycles of the $T^{3}$ we are compactifying on.
Secondly, the potential in the SS theory comes with a factor
$1/\Im{\rm m}(\tau)$ where $\tau$ is the complex scalar field in the
$Sl(2,\mathbb{R})/SO(2)$ coset while in our case the potential carries
a factor $|\tau|^{2}|/\Im{\rm m}(\tau)$. This factor is precisely the
S~dual of the one in SS ({\em i.e.}~they are related by the
$Sl(2,\mathbb{R})$ transformation $\tau\rightarrow -1/\tau$) in the
ungauged, massless theory and we will see that the KK vectors are also
the S~dual of the vectors coming from the 11-dimensional 3-form (also
in the ungauged, massless, theory because the gauging explicitly
breaks S~duality).

It is clear, then, that if we wanted to obtain gauged 8-dimensional
supergravity from the ungauged theory by the standard gauging
procedure, we could have chosen to gauge the KK vectors and we would
have obtained the SS theory or we could have chosen to gauge the
vectors associated to membranes wrapped on 2 cycles and we would have
obtained our theory. Both theories are related by an S~duality
transformation {\it which is not a symmetry}, neither of the theory
nor of the equations of motion but related two different theories. The
S~duality transformation can also be understood as a field
redefinition.  In the first case we could have explained its
11-dimensional origin in more or less standard terms but in the second
only by appealing to the ideas of Refs.~\cite{kn:BLO,kn:MO} we could
get some 11-dimensional understanding.  This is essentially our main
result.

The rest of the paper is organized as follows: in
Section~\ref{sec-masslesssugra} we perform the dimensional reduction
of 11-dimensional supergravity to $d=8$. In
Section~\ref{sec-massivesugra} we perform the dimensional reduction of
{\it massive} 11-dimensional supergravity to $d=8$ dimensions to
obtain an $Sl(3,\mathbb{R})$ multiplet of massive $d=8$ supergravity
theories with $SO(3)$ gauge symmetry.  In Section~\ref{sec-vacua} we
study the vacua of this theory.  In Section~\ref{sec-conclusions} we
present our conclusions, comment on possible interpretations and
future directions of work.

%%%%%%%%%%%%%%%%%%%%%%%%%%%%%%%%%%%%%%%%%%%%%%%%%%%%%%%%%%%%%%%%%%%%%%%%%%
\section{Direct dimensional reduction of $D=11$ Supergravity on $T^{3}$}
%%%%%%%%%%%%%%%%%%%%%%%%%%%%%%%%%%%%%%%%%%%%%%%%%%%%%%%%%%%%%%%%%%%%%%%%%%
\label{sec-masslesssugra}

Our goal in this section is to perform the standard dimensional
reduction of 11-dimensional supergravity to obtain the theory
describing the massless fields that arise in the compactification of
that theory on $T^{3}$. 11-dimensional supergravity was compactified
to 8 dimensions by Salam and Sezgin in Ref.~\cite{kn:SaSe}, but they
used Scherk and Schwarz's GDR \cite{kn:SS} to compactify on an $SU(2)$
internal manifold to obtain a $SU(2)$ gauged $d=8$ supergravity. The
gauged theory does not exhibit all the duality symmetries of the
ungauged one. In particular, gauging usually break all
electric-magnetic dualities. Since we are interested in the theory
with all its dualities and its corresponding massive version, we start
by the standard dimensional reduction of 11-dimensional supergravity.

%%%%%%%%%%%%%%%%%%%%%%%%%%%%%%%%%%%%%%%%%%%%%%%%%%%%%%%%%%%%%%%%%%%%%%%%%%%
\subsection{$d=11$ Supergravity}
%%%%%%%%%%%%%%%%%%%%%%%%%%%%%%%%%%%%%%%%%%%%%%%%%%%%%%%%%%%%%%%%%%%%%%%%%%%

The bosonic fields of $N=1,d=11$ supergravity are the Elfbein and a
3-form potential\footnote{Index conventions: $\hat{\hat{\mu}}$
  ($\hat{\hat{a}}$) are curved (flat) 11-dimensional, $\mu$ ($a$) are
  curved (flat) 8-dimensional, and $m$ ($i$) are curved (flat)
  3-dimensional (compact space). Our signature is $(+-\cdots-)$.}

\begin{equation}
\left\{\hat{\hat{e} \,}_{\hat{\hat{\mu}}}{}^{\hat{\hat{a}}},
\hat{\hat{C} \,}_{\hat{\hat{\mu}}\hat{\hat{\nu}}\hat{\hat{\rho}}}
\right\}\, .
\end{equation}

\noindent The field strength of the 3-form is

\begin{equation}
\hat{\hat{G} \,} = 4\, \partial\, \hat{\hat{C} \,}\, ,
\end{equation}

\noindent and is obviously invariant under the gauge transformations

\begin{equation}
\label{eq:3formgauge}
\delta\, \hat{\hat{C} \,}= 3\, \partial\, \hat{\hat{\chi}}\, ,
\end{equation}

\noindent where $\hat{\hat{\chi}}$ is a 2-form. 

\noindent The action for these bosonic fields is

\begin{equation}
\label{eq:action11}
 \hat{\hat{S} \,} = 
\int d^{11} \hat{\hat{x}} \sqrt{|\hat{\hat{g}}|}
 \left[ \hat{\hat{R} \,} -{\textstyle\frac{1}{2 \cdot 4!}}
\hat{\hat{G} \,}{}^{2}
 -{\textstyle\frac{1}{6^{4}}}
{\textstyle\frac{1}{\sqrt{|\hat{\hat{g}}|}}}\, \hat{\hat{\epsilon} \,}
\partial\hat{\hat{C} \,}\partial\hat{\hat{C} \,}\hat{\hat{C} \,} \right]\, .
\end{equation}

%%%%%%%%%%%%%%%%%%%%%%%%%%%%%%%%%%%%%%%%%%%%%%%%%%%%%%%%%%%%%%%%%%%%%%
\subsection{Dimensional Reduction}
%%%%%%%%%%%%%%%%%%%%%%%%%%%%%%%%%%%%%%%%%%%%%%%%%%%%%%%%%%%%%%%%%%%%%%

\noindent The KK Ansatz for the Elfbein is

\begin{equation}
\left( \hat{\hat{e} \,}_{\hat{\hat{\mu}}}{}^{\hat{\hat{a}}} \right) = 
\left(
\begin{array}{cr}
e_{\mu}{}^{a} & e_{m}{}^{i}A^{m}{}_{\mu} \\
&\\
0             & e_{m}{}^{i}                \\
\end{array}
\right)
\, , 
\hspace{1cm}
\left(\hat{\hat{e} \,}_{\hat{\hat{a}}}{}^{\hat{\hat{\mu}}} \right) =
\left(
\begin{array}{cr}
e_{a}{}^{\mu} & -A^{m}{}_{a}   \\
& \\
0             &  e_{i}{}^{m}  \\
\end{array}
\right)\, , 
\label{eq:elfbein}
\end{equation}

\noindent where $A^{m}{}_{a}=e_{a}{}^{\mu}A^{m}{}_{\mu}$. We define the
internal metric on $T^{3}$ by

\begin{equation}
G_{mn}=e_{m}{}^{i}e_{n\, j}=-e_{m}{}^{i}e_{n}{}^{j}\delta_{ij}\, .
\end{equation}

\noindent Under global transformations in the internal space

\begin{equation}
x^{m\ \prime} = \left(R^{-1\ T}\right)^{m}{}_{n}\ x^{n} +a^{m}\, ,
\hspace{1cm}
R\in GL(3,\mathbb{R})\, ,
\end{equation}

\noindent objects with internal space indices 
(the internal metric $G=(G_{mn})$ and the KK vectors
$\vec{A}_{\mu}=(A^{m}{}_{\mu})$) transform as follows:

\begin{equation}
G^{\prime}= RGR^{T}\, ,  
\hspace{1cm}
\vec{A}^{\prime}_{\mu} = (R^{-1})^{T} \vec{A}_{\mu}\, .
\end{equation}

\noindent We know that $GL(3,\mathbb{R})$ can be decomposed in
$Sl(3,\mathbb{R})\times \mathbb{R}^{+}\times\mathbb{Z}_{2}$ and any
matrix $R$, forgetting its $\mathbb{Z}_{2}$ part
(we will focus on $GL(3,\mathbb{R})/\mathbb{Z}_{2}\sim
Sl(3,\mathbb{R})\times \mathbb{R}^{+}$), can therefore be
decomposed into

\begin{equation}
R= c\Lambda \, ,
\hspace{.5cm}
\Lambda \in Sl(3,\mathbb{R})\, ,
\hspace{.5cm}
c\in \mathbb{R}^{+}\, .
\end{equation}

\noindent We want to separate fields that transform under the 
different factors.  First we define the symmetric $Sl(3,\mathbb{R})$
matrix

\begin{equation}
{\cal M} = -G/| {\rm det}\ G |^{1/3}\, ,
\label{eq:defM}\end{equation}

\noindent and the scalar

\begin{equation}
\sqrt{|{\rm det}\ G|}=e^{-\varphi}\, .
\end{equation}

\noindent Now, under $Sl(3,\mathbb{R})$ only ${\cal M}$ and $\vec{A}_{\mu}$
transform:

\begin{equation}
{\cal M}^{\prime} = \Lambda {\cal M} \Lambda^{T}\, ,  
\hspace{.5cm}
\vec{A}_{\mu}^{\prime} = (\Lambda^{-1})^{T}\vec{A}_{\mu}\, ,
\end{equation}

\noindent that is, $\vec{A}_{\mu}$ transforms contravariantly,
while under $\mathbb{R}^{+}$ rescalings only $\varphi$ and
$\vec{A}_{\mu}$ transform:

\begin{equation}
\varphi^{\prime} = \varphi -\log{c}\, ,
\hspace{.5cm}
\vec{A}_{\mu}^{\prime} = c \vec{A}_{\mu}\, .
\end{equation}

\noindent For future convenience, we label the KK vector with an upper 
index $1$, {\em i.e.}~$A^{1\, m}{}_{\mu}$.

Using the standard techniques, the above Elfbein Ansatz, and rescaling
the resulting 8-dimensional metric to the Einstein frame

\begin{equation}
\label{eq:masscale}
g_{\mu\nu} =e^{\varphi/3}g_{E\ \mu\nu}\, ,  
\end{equation}

\noindent one finds

\begin{equation}
\label{eq:curvature}
\begin{array}{rcl}
{\displaystyle\int} 
d^{11}\hat{\hat{x}}\, \sqrt{|\hat{\hat{g}}|}\ 
\left[\ \hat{\hat{R} \,}\ \right] & = & 
{\displaystyle\int} d^{8}x\, \sqrt{|g_{E}|}\
\left[
R_{E} +\frac{1}{2}\left(\partial\varphi \right)^{2}
\right.\\
& & \\
& & 
\left.
+{\textstyle\frac{1}{4}} {\rm Tr} 
\left(\partial {\cal M} {\cal M}^{-1}\right)^{2} 
-{\textstyle\frac{1}{4}} e^{-\varphi} 
F^{1\, m}{\cal M}_{mn}F^{1\, n}
\right]\, , 
\end{array}
\end{equation}

\noindent where 

\begin{equation}
\label{eq:fieldstrength1}
F^{1\, m} = 2\partial A^{1\, m}\, .  
\end{equation}

The kinetic term for ${\cal M}$ is just an $Sl(3,\mathbb{R})/SO(3)$
sigma model.

\noindent The fields arising from 
$\hat{\hat{C} \,}_{\hat{\hat{\mu}}\hat{\hat{\nu}}\hat{\hat{\rho}}}$
are $\left\{ C_{\mu\nu\rho}, B_{\mu\nu\, m}, A^{2\, m}{}_{\mu}, a
\right\}$.  We decompose the 11-dimensional 3-form by identifying objects
with flat 11- and 8-dimensional flat indices (up to factors coming
from the rescaling of the metric) as

\begin{equation}
\label{eq:3formpotentialflat}
\begin{array}{rcl}
&& \hat{\hat{C} \,}_{abc} = e^{-\varphi/2} C_{abc}\, ,\\ 
 & & \\
&& \hat{\hat{C} \,}_{ab\, m} = 
e^{-\varphi/3}B_{m\, ab}\, ,\\
 & & \\
&& \hat{\hat{C} \,}_{a\, mn} = 
\epsilon_{mnp} e^{-\varphi/6}A^{2\, p}{}_{a}\, ,\\ 
\end{array}
\end{equation}

\noindent which implies, for curved components

\begin{equation}
\label{eq:3formpotential}
\begin{array}{rcl}
&& \hat{\hat{C} \,}_{\mu\nu\rho} = 
C_{\mu\nu\rho} 
+3A^{1\, m}{}_{[\mu}B_{|m|\,\nu\rho]} 
+3\epsilon_{mnp} A^{1\, m}{}_{[\mu} A^{1\, n}{}_{\nu} A^{2\, p}{}_{\rho]}
+\epsilon_{mnp}\, a\, 
A^{1\, m}{}_{[\mu} A^{1\, n}{}_{\nu} A^{1\, p}{}_{\rho]}\, ,\\
 & & \\
&& \hat{\hat{C} \,}_{\mu\nu\, m} = 
B_{m\, \mu\nu}
+2\epsilon_{mnp} A^{1\, n}{}_{[\mu} A^{2\, p}{}_{\nu]}
+\epsilon_{mnp}\, a\, A^{1\, n}{}_{[\mu} A^{1\, p}{}_{\nu]}\, ,\\
 & & \\
&& \hat{\hat{C} \,}_{\mu\, mn} = 
\epsilon_{mnp} A^{2\, p}{}_{\mu} 
+\epsilon_{mnp} a\, A^{1\, p}{}_{\mu}\, , \\
 & & \\
&& \hat{\hat{C} \,}_{mnp} = \epsilon_{mnp}\, a\, .\\
 & & \\
\end{array}
\end{equation}

\noindent These fields inherit the following gauge transformations from
11-dimensional gauge and general coordinate transformations of
$\hat{\hat{C}\,}$:

\begin{equation}
\begin{array}{rcl}
& & \delta C = 3\partial\Lambda -6A^{1\, m}\partial \Lambda_{m}
+3\epsilon_{mnp} A^{1\, m} A^{1\, n} \partial\Lambda^{2\, p}\, ,\\
& & \\
& & \delta B_{m} = 2 \partial \Lambda_{m} 
-2\epsilon_{mnp}A^{1\, n} \partial\Lambda^{2\, p}\, ,\\
& & \\
& & \delta A^{2\, m} = \partial\Lambda^{2\, m}\, .\\
\end{array}
\end{equation}

\noindent In particular we see that this choice implies that these potentials
do not transform under reparametrizations of the internal torus
$\delta A^{1\, m}=\partial \Lambda^{1\, m}$.  The gauge-invariant
field strengths of the above fields are

\begin{equation}
\label{eq:fieldstrengths2}
\begin{array}{rcl}
& & G = 4 \partial C +6 F^{1\, m} B_{m}\, ,\\
& & \\
& & H_{m} = 3\partial B_{m} +3\epsilon_{mnp}F^{1\, n}A^{2\, p}\, ,\\
& & \\
& & F^{2\, m} = 2\partial A^{2\, m}\, ,\\
\end{array}
\end{equation}

\noindent and lead to the following non-trivial Bianchi identities:

\begin{equation}
\begin{array}{rcl}
& & \partial G = 2 F^{1\, m} H_{m}\, ,\\
& & \\
& & \partial H_{m} = \frac{3}{2}\epsilon_{mnp}F^{1\, n}F^{2\, p}\, ,\\
\end{array}
\end{equation}

Using now

\begin{equation}
 e_{i}{}^{m}\, e_{j}{}^{n}\, e_{k}{}^{p}\, \epsilon_{mnp}
 = \det(e^{-1})\, \epsilon_{ijk}\, ,
\end{equation}

\noindent with $\det(e^{-1})=e^{\varphi}$, we get the following 
decomposition of the 11-dimensional 4-form field strength
into the above 8-dimensional field strengths:

\begin{equation}
\label{eq:3formfieldstrength}
\begin{array}{rcl}
&& \hat{\hat{G} \,}_{abcd} = 
e^{-2\varphi/3} G_{abcd}\, ,\\
 & & \\
&& \hat{\hat{G} \,}_{abc\, i} = 
e^{-\varphi/2} e_{i}{}^{m}\, H_{m\, abc}\, , \\
 & & \\
&& \hat{\hat{G} \,}_{ab\, ij} = 
e^{2\varphi/3}\, \epsilon_{ijk}\, e_{p}{}^{k}\,
\left[ F^{2\, p}{}_{ab} + a\, F^{1\, p}{}_{ab} \right]\, , \\
 & & \\
&& \hat{\hat{G} \,}_{a\, ijk} = 
e^{5\varphi/6}\, \epsilon_{ijk}\, \partial_{a}\, a\, .\\
 & & \\
\end{array}
\end{equation}

We can now reduce the kinetic term of the 11-dimensional 3-form. The
result can be combined with the result of the reduction of the
Einstein-Hilbert term, giving

\begin{equation}
\begin{array}{rcl}
{\displaystyle\int} d^{11} \hat{\hat{x}} \sqrt{|\hat{\hat{g}}|}
 \left[ \hat{\hat{R} \,} -\frac{1}{2 \cdot 4!}\hat{\hat{G}\,}{}^{2} \right]
& = & {\displaystyle\int} d^{8}x \sqrt{|g_{E}|}\,
\left[ 
R_{E} 
+\frac{1}{4}{\rm Tr}\left(\partial {\cal M}{\cal M}^{-1}\right)^{2}
+\frac{1}{4}{\rm Tr}\left(\partial {\cal W}{\cal W}^{-1}\right)^{2}
 \right.\\
& & \\
& & 
\left.
-\frac{1}{4}F^{i\, m}{\cal M}_{mn}{\cal W}_{ij} F^{j\, n}
+\frac{1}{2\cdot 3!} H_{m}{\cal M}^{mn} H_{n}
-\frac{1}{2\cdot 4!} e^{-\varphi} G^{2} \right]\, ,\\
 \end{array}
\end{equation}

\noindent where we have introduced the symmetric 
$Sl(2,\mathbb{R})/SO(2)$ matrix

\begin{equation}
{\cal W}
=\frac{1}{\Im{\rm m}(\tau)}
\left(
\begin{array}{cc}
|\tau|^{2} &    \Re{\rm e}(\tau)  \\
& \\
\Re{\rm e}(\tau)    &  1          \\
\end{array}
\right)\, ,
\end{equation}

\noindent whet $\tau$ is the standard complex combination

\begin{equation}
\tau= a+ie^{-\varphi}\, .  
\end{equation}
 
Let us now reduce the Chern-Simons term also using tangent space indices
and taking into account the definition, in any dimension

\begin{equation}
\epsilon^{\mu_{1}\cdots\mu_{d}} = \sqrt{|g|}\, \epsilon^{a_{1}\cdots a_{d}} 
e_{a_{1}}{}^{\mu_{1}} \ldots e_{a_{d}}{}^{\mu_{d}} \, .
\end{equation}

The final result is\footnote{One of our coefficients differs from the
  corresponding in Ref.~\cite{kn:SaSe}. Our Chern-Simons term has been
  explicitly checked to be gauge-invariant.}

\begin{equation}
\label{eq:d8masslessaction}
  \begin{array}{rcl}
S & = & {\displaystyle\int} d^{8}x \sqrt{|g_{E}|}\,
\left\{ 
R_{E} 
+\frac{1}{4}{\rm Tr}\left(\partial {\cal M}{\cal M}^{-1}\right)^{2}
+\frac{1}{4}{\rm Tr}\left(\partial {\cal W}{\cal W}^{-1}\right)^{2}
 \right.\\
& & \\
& & 
-\frac{1}{4}F^{i\, m}{\cal M}_{mn}{\cal W}_{ij} F^{j\, n}
+\frac{1}{2\cdot 3!} H_{m}{\cal M}^{mn} H_{n}
-\frac{1}{2\cdot 4!} e^{-\varphi} G^{2} \, ,\\
& & \\
& & 
-\frac{1}{6^{3}\cdot 2^{4}}
{\textstyle\frac{1}{\sqrt{|g_{E}|}}}\, \epsilon
\left[GGa -8G H_{m} A^{2\, m} +12 G(F^{2\, m}+aF^{1\, m})B_{m} \right.\\
&& \\
& & 
\left.\left.
-8\epsilon^{mnp} H_{m}H_{n}B_{p} -8G\partial a C 
-16 H_{m}(F^{2\, m}+aF^{1\, m}) C \right]\right\} \, .\\
\end{array}
\end{equation}

The kinetic terms (except for that of $C$) are explicitly invariant under
$Sl(2,\mathbb{R})$  transformations

\begin{equation}
\label{eq:sl2rules1}
{\cal W}^{\prime} = \Lambda {\cal W} \Lambda^{T}\, ,\\
\hspace{1cm}
F^{i\, m\, \prime} = F^{j\, m}\left( \Lambda^{-1} \right)_{j}{}^{i}\, ,
\hspace{1cm}
\Lambda \in Sl(2,\mathbb{R})\, ,
\end{equation}

\noindent and $Sl(3,\mathbb{R})$ transformations

\begin{equation}
{\cal M}^{\prime} = K {\cal M} K^{T}\, ,\\
\hspace{.5cm}
F^{i\, m\, \prime} = F^{i\, n}\left( K^{-1} \right)_{n}{}^{m}\, ,
\hspace{.5cm}
H^{\prime}_{m} = K_{m}{}^{n} H_{n} \, ,
\hspace{.5cm}
K \in Sl(3,\mathbb{R})\, .
\end{equation}

\noindent The kinetic term of $C$ and the Chern-Simons term are 
not invariant as a matter of fact. However, let us look into the
equations of motion of $C$. We can write them, together with the
Bianchi identity, in the following form:

\begin{equation}
\partial G^{i} =2 F^{i\, m}H_{m}\, ,  
\end{equation}

\noindent where 

\begin{equation}
\label{eq:sl2rules2}
G^{1} \equiv G\, ,
\hspace{1cm}
G^{2} \equiv -e^{-\varphi}{}^{\star} G-a G\, . 
\end{equation}

$G^{i}$ transforms as a doublet under $Sl(2,\mathbb{R})$ (just like
the doublet $F^{i\, m}$) and therefore, the above equation of motion
is covariant under $Sl(2,\mathbb{R})$ electric-magnetic duality
transformations. The remaining equations of motion are covariant under
$Sl(2,\mathbb{R})$ transformations as well. The structures are very
similar to those of $N=4,d=4$ supergravity (see {\em
  e.g.}~Ref.~\cite{kn:L-TO1}), the obvious difference being that in
four dimensions we dualize 2-form field strengths and in eight
dimensions we dualize 4-form field strengths. This duality was first
described in Ref.~\cite{kn:ILPT} and is part of a series of
electric-magnetic dualities present in type~II theories in any
dimension (the 6-dimensional version was studied in Ref.~\cite{kn:BBO}
and a general discussion can be found in Ref.~\cite{kn:L-TO2}).

Let us summarize our results: the 8-dimensional supergravity theory we
have just obtained has, then, the bosonic fields

\begin{equation}
\{g_{\mu\nu}, C, B_{m}, A^{1\, m}, A^{2\, m}, a,\varphi,{\cal M}_{mn} \}\, ,
\end{equation}

\noindent with field strengths given by Eqs.~(\ref{eq:fieldstrength1},
\ref{eq:fieldstrengths2}) and action given by
Eq.~(\ref{eq:d8masslessaction}). The scalars parametrize
$Sl(3,\mathbb{R})/SO(3)$ and $Sl(2,\mathbb{R})/SO(2)$ sigma models.
The action has the global invariance group $Sl(3,\mathbb{R})$ but the
equations of motion are also invariant under $Sl(2,\mathbb{R})$
electric-magnetic duality transformations.

%%%%%%%%%%%%%%%%%%%%%%%%%%%%%%%%%%%%%%%%%%%%%%%%%%%%%%%%%%%%%%%%%%%%%%
\section{Dimensional Reduction of Massive 11-Dimensional Supergravity}
%%%%%%%%%%%%%%%%%%%%%%%%%%%%%%%%%%%%%%%%%%%%%%%%%%%%%%%%%%%%%%%%%%%%%%
\label{sec-massivesugra}

In this section we are going to perform the dimensional reduction to
eight dimensions of a further generalization of the massive
11-dimensional theory proposed in Refs.~\cite{kn:BLO,kn:MO}. We are
going to use the same field definitions as in the previous section
since we want to recover that theory in the massless limit. First we
start by describing the massive 11-dimensional theory.

%%%%%%%%%%%%%%%%%%%%%%%%%%%%%%%%%%%%%%%%%%%%%%%%%%%%%%%%%%%%%%%%%%%%%%
\subsection{Massive 11-Dimensional Supergravity}
%%%%%%%%%%%%%%%%%%%%%%%%%%%%%%%%%%%%%%%%%%%%%%%%%%%%%%%%%%%%%%%%%%%%%%

The theory we are considering has three Killing vectors
$\hat{\hat{k}}_{(n)}$, which are defined by

\begin{equation}
\hat{\hat{k}}_{(m)}{}^{\hat{\hat{\mu}}}
\partial_{\hat{\hat{\mu}}}= \partial_{m} 
\hspace{.5cm}\rightarrow\hspace{.5cm}
\hat{\hat{k}}_{(m)}{}^{\hat{\hat{\mu}}}
\hat{\hat{k}}_{(n)}{}^{\hat{\hat{\nu}}}
\hat{\hat{g}}_{\hat{\hat{\mu}}\hat{\hat{\nu}}}=
G_{nm}\; ,
\label{eq:KillDef}
\end{equation}

\noindent and a symmetric matrix $Q^{mn}$, which we will leave 
arbitrary for the moment. With these elements and the 2-form gauge
parameter $\hat{\hat{\chi}}$ we construct the massive gauge
parameter\footnote{$i_{\hat{\hat{k}}}\hat{\hat{T}\,}$ stands for the
  contraction of the last index of the tensor $\hat{\hat{T}\,}$ with
  the vector $\hat{\hat{k}}$.}

\begin{equation}
\hat{\hat{\lambda}}{}^{(n)} 
\equiv
-i_{\hat{\hat{k}}_{\scriptscriptstyle (m)}}\hat{\hat{\chi}} 
Q^{nm} \, ,
\end{equation}

\noindent and define the massive gauge transformations of the two 
11-dimensional fields:

\begin{equation}
\left\{\begin{array}{rcl}
\delta_{\hat{\hat{\chi}}}
\hat{\hat{g}}_{\hat{\hat{\mu}}\hat{\hat{\nu}}} & = &
2\hat{\hat{\lambda}}{}^{(n)}{}_{(\hat{\hat{\mu}}}\,  
\hat{\hat{k}}_{(n)}{}^{\hat{\hat{\rho}}}
\hat{\hat{g}}_{\hat{\hat{\nu}})\hat{\hat{\rho}}} \; ,\\
& & \\
\delta_{\hat{\hat{\chi}}} \hat{\hat{C}\,} & = & 
3\partial\hat{\hat{\chi}} 
+3\hat{\hat{\lambda}}{}^{(n)}
\left(i_{\hat{\hat{k}}_{\scriptscriptstyle (n)}}\hat{\hat{C}\,}\right) \; .\\
\end{array}
\right.
\end{equation}

The 4-form field strength is given by 

\begin{equation}
\hat{\hat{G}\,} = 
4\partial \hat{\hat{C}\,} 
+3\left(i_{\hat{\hat{k}}_{\scriptscriptstyle (m)}}\hat{\hat{C}\,}\right)Q^{mn}
\left(i_{\hat{\hat{k}}_{\scriptscriptstyle (n)}}\hat{\hat{C}\,}\right)\, .
\end{equation}

\noindent The proposed 11-dimensional massive supergravity then 
reads\footnote{This action is the same as in Ref.~\cite{kn:MO} but
  here we have taken the contorsion part out of the connection used in
  the curvature.}

\begin{equation}
\begin{array}{rcl}
\hat{\hat{S}\,}
& = &
{\displaystyle\int} d^{11}\hat{\hat{x}}\sqrt{|\hat{\hat{g}}|}\,
\left\{
\hat{\hat{R}\,}\left(\hat{\hat{g}}\right)
+\textstyle{\frac{1}{2}}\left(d\hat{\hat{k}}_{(n)}\right)
   {}_{\hat{\hat{\mu}}\hat{\hat{\nu}}}Q^{nm}
   \left( i_{\hat{\hat{k}}_{(m)}}\hat{\hat{C}\,}\right)
   {}^{\hat{\hat{\mu}}\hat{\hat{\nu}}}
-\textstyle{\frac{1}{2\cdot 4!}} \hat{\hat{G}\,}{}^{2}
\right.\\
& & \\
& &
-\hat{\hat{K}}_{\hat{\hat{\mu}}\hat{\hat{\nu}}\hat{\hat{\rho}}}
\hat{\hat{K}}{}^{\hat{\hat{\nu}}\hat{\hat{\rho}}\hat{\hat{\mu}}}
+\textstyle{\frac{1}{2}}
\left( \hat{\hat{k}}_{(n)\ \hat{\hat{\mu}}}Q^{nm}
\hat{\hat{k}}_{(m)}{}^{\hat{\hat{\mu}}}\right)^{2}
-\left(  \hat{\hat{k}}_{(n)\ \hat{\hat{\mu}}}Q^{nm}
\hat{\hat{k}}_{(m)\ \hat{\hat{\nu}}}\right)^{2}
\\
& & \\
& &
-\frac{1}{6^{4}}
\frac{\hat{\hat{\epsilon}}}{\sqrt{|\hat{\hat{g}}|}}
\left\{
\partial\hat{\hat{C}\,}\partial\hat{\hat{C}\,}\hat{\hat{C}\,}
+ \frac{9}{8}\partial\hat{\hat{C}\,}\hat{\hat{C}\,}
\left(i_{\hat{\hat{k}}_{(n)}}\hat{\hat{C}\,}\right)
Q^{nm}
\left(i_{\hat{\hat{k}}_{(m)}}\hat{\hat{C}\,}\right)
\right. \\
& & \\
& &
\left.
\left.
+ \frac{27}{80}
\hat{\hat{C}\,}
\left[
\left(i_{\hat{\hat{k}}_{(n)}}\hat{\hat{C}\,}\right)
 Q^{nm}
\left(i_{\hat{\hat{k}}_{(m)}}\hat{\hat{C}\,}\right)
\right]^{2}
\right\}
\right\}\; ,
\end{array}
\label{eq:masact.11}
\end{equation}

\noindent where we have defined a {\em contorsion} tensor

\begin{equation}
\hat{\hat{K}}_{\hat{\hat{a}}\hat{\hat{b}}\hat{\hat{c}}}
\;=\; {\textstyle\frac{1}{2}}
\left(
\hat{\hat{T}\,}_{\hat{\hat{a}}\hat{\hat{c}}\hat{\hat{b}}}
\; +\; \hat{\hat{T}\,}_{\hat{\hat{b}}\hat{\hat{c}}\hat{\hat{a}}}
\; -\; \hat{\hat{T}\,}_{\hat{\hat{a}}\hat{\hat{b}}\hat{\hat{c}}}
\right) \; ,
\label{eq:torsion.2}
\end{equation}

\noindent and where the {\em torsion} is defined by

\begin{equation}
\hat{\hat{T}\,}_{\hat{\hat{\mu}}\hat{\hat{\nu}}}{}^{\hat{\hat{\rho}}}
= -\left(i_{\hat{\hat{k}}_{(n)}}\hat{\hat{C}\,}
\right)_{\hat{\hat{\mu}}\hat{\hat{\nu}}} Q^{nm}
\hat{\hat{k}}_{(m)}{}^{\hat{\hat{\rho}}}\, .
\end{equation}

Note that when considering the above theory defined with $n$ Killing
vectors, we should take $Q$ to be invertible, since otherwise we could
diagonalize $Q$ and end up with a theory defined with less Killing
vectors. {\em E.g.} in the case at hand, taking $Q$ to have 2 zero
eigenvalues, would lead to Romans' theory compactified over a 2-torus.

%%%%%%%%%%%%%%%%%%%%%%%%%%%%%%%%%%%%%%%%%%%%%%%%%%%%%%%%%%%%%%%%%%%%%%
\subsection{Dimensional Reduction}
%%%%%%%%%%%%%%%%%%%%%%%%%%%%%%%%%%%%%%%%%%%%%%%%%%%%%%%%%%%%%%%%%%%%%%

We start by reducing the massive gauge parameters, the massive gauge
transformations of the fields (whose definitions are the same as in
the massless case) and the field strengths.  We first make the
following definitions inspired by $SO(3)$ gauge theory (see
Appendix~\ref{sec-so3}) of which $A^{2\, m}$ is going to play the role
of gauge field:

\begin{equation}
f_{mn}{}^{p}\equiv\epsilon_{mnq}Q^{qp}\, ,
\hspace{1cm}
\sigma^{m}{}_{n} = \Lambda^{2\, p}f_{pn}{}^{m}\, .
\end{equation}

\noindent We are also going to use the definitions of $SO(3)$
gauge covariant derivative that can be found in that appendix.  We get

\begin{equation}
\left\{
\begin{array}{rcl}
\hat{\hat{\lambda}}{}^{(m)}{}_{\mu} & = & -\Lambda_{n\, \mu}Q^{nm}\, ,\\
& & \\
\hat{\hat{\lambda}}{}^{(m)}{}_{n} & = & \sigma^{m}{}_{n}\, .\\
\end{array}
\right.
\end{equation}

The massive gauge transformations are

\begin{equation}
\begin{array}{rcl}
\delta {\cal M}_{mn} & = & -{\cal M}_{pn}\sigma^{p}{}_{m}
-{\cal M}_{mp}\sigma^{p}{}_{n}\, ,\\
& & \\
\delta A^{1\, m} & = & \sigma^{m}{}_{n}A^{1\, n} -Q^{mn}\Lambda_{n}\, ,\\
& & \\
\delta A^{2\, m} & = & {\cal D}\Lambda^{2\, m}\, ,\\
& & \\
\delta B_{m} & = & 2\partial\Lambda_{m} -B_{n}\sigma^{n}{}_{m} 
+2 \partial\sigma^{q}{}_{m}(Q^{-1})_{qn} A^{1\, n}\, ,\\
& & \\
\delta C & = & \, 3\partial\Lambda_{(2)}
     \,-\, 3\partial\sigma^{m}{}_{n}\left( Q^{-1}\right)_{mp}A^{1\ n}A^{1\ p}
          \,-\,6\Lambda_{m}\partial A^{1\ m}.\\
\end{array}
\end{equation}

We see that there are two kinds of massive gauge transformations:
those generated by the 1-form parameters $\Lambda_{m}$ which are
standard massive gauge transformations that shift the vectors $A^{1\,
  m}$ so they can be completely gauged away, and those generated by
the scalar parameters $\Lambda^{2\, m}$ that take the form of $SO(3)$
gauge transformations in a non-standard basis (see
Appendix~\ref{sec-so3}).

The gauge-covariant field strengths are

\begin{equation}
\label{eq:massivefieldstrengths}
\begin{array}{rcl}
& & G = 4 \partial C +6 F^{1\, m} B_{m}-3B_{m}Q^{mn}B_{n}\, ,\\
& & \\
& & H_{m} = 3\partial B_{m} +3\epsilon_{mnp}F^{1\, n}A^{2\, p}\, ,\\
& & \\
& & F^{2\, m} = 2\partial A^{2\, m}-f_{pq}{}^{m}A^{2\, p}A^{2\, q}
\, ,\\
& &\\
& & F^{1\, m} = 2\partial A^{1\, m}+Q^{mn}B_{n}\, ,\\
& & \\
& & {\cal D}{\cal M}_{mn} = 
\partial {\cal M}_{mn}+f_{pm}{}^{q}A^{2\, p}{\cal M}_{qn}
+f_{pn}{}^{q}A^{2\, p}{\cal M}_{mq}\, ,\\
\end{array}
\end{equation}

\noindent where the $F^{1\, m}$ appearing in $G$ and $H_{m}$ is the 
massive one.

The field strength of $A^{2\, m}$ is just an $SO(3)$ gauge field
strength while the field strength of the scalar matrix ${\cal M}_{mn}$
is the $SO(3)$ gauge covariant derivative of an object with two
covariant vector indices. The field strength of $A^{1\, m}$ has a
different form, typical of massive theories and it is ready to give
mass terms (a mass matrix) for the 2-forms $B_{m}$ when the $A^{1\,
  m}$s which are nothing but St\"uckelberg fields for the $B_{m}$s,
are gauged away. The field strength of $B_{m}$ could also be written
in this way

\begin{equation}
H_{m} = 3{\cal D} B_{m} +6\epsilon_{mnp}\partial A^{1\, n}A^{2\, p}\, .
\end{equation}

\noindent The first term is the $SO(3)$ gauge covariant derivative
of an $SO(3)$-covariant object (antisymmetrized, as usual in all three
lower indices). The second term restores the invariance under
$\Lambda_{m}$ transformations which is broken in the covariant
derivative term.

Thus, all field strengths are invariant under $\Lambda_{m}$
transformations and covariant under $\Lambda^{2\, m}$ ({\em i.e.}
$SO(3)$) transformations

\begin{equation}
\begin{array}{rcl}
& & \delta H_{m} =  -H_{n}\sigma^{n}{}_{m}\, ,\\
& & \\
& & \delta F^{1,2\, m} = \sigma^{m}{}_{n} F^{1,2\, n}\, ,\\
\end{array}
\end{equation}

\noindent except for $G$, which is invariant.

Let us have a look at the part of the 11-dimensional massive theory
that, naively, will lead to the potential for the
$Sl(3,\mathbb{R})/SO(3)$ scalar fields.  This term is easily found to
be

\begin{equation}
  \begin{array}{rcl}
\sqrt{|\hat{\hat{g}}|}\,\left[
  \textstyle{\frac{1}{2}}
\left(\hat{\hat{k}}_{(n)\, \hat{\hat{\mu}}}
Q^{nm} \hat{\hat{k}}_{(m)}{}^{\hat{\hat{\mu}}}\right)^{2}
-\left(\hat{\hat{k}}_{(n)\, \hat{\hat{\mu}}}
Q^{nm}\hat{\hat{k}}_{(m)\, \hat{\hat{\nu}}}\right)^{2}
\right]= 
& & \\
& & \\
& & 
\hspace{-6cm}
=\sqrt{|g_{E}|}\, \left\{ 
\textstyle{\frac{1}{2}}e^{-\varphi}
\left[{\rm Tr}\,\left( Q\mathcal{M}\right)
    \right]^{2}
    \,-\, 
   2{\rm Tr}\, \left( 
          \mathcal{M}Q\mathcal{M}Q
      \right)
  \right\}\, .\\
\end{array}
\label{eq:Potential1}
\end{equation}

The two terms in the eleven dimensional Lagrangian that are modifying
the field strengths of the objects coming from the metric are,

\begin{equation}
\label{eq:yoquese}
\begin{array}{rcl}
\textstyle{\frac{1}{2}}
\sqrt{|\hat{\hat{g}}|}\, 
\left(d\hat{\hat{k}}_{(n)}\right)_{\hat{\hat{\mu}}\hat{\hat{\nu}}}
Q^{nm} \left(i_{\hat{\hat{k}}_{(m)}}\hat{\hat{C}\,}
\right)^{\hat{\hat{\mu}}\hat{\hat{\nu}}}= & & \\
& & \\
& & \hspace{-7cm}
=\sqrt{|g_{E}|}\, 
\left\{
-\textstyle{\frac{1}{2}} e^{-\varphi} \left( dA^{1\ n}\right)_{\mu\nu}
   \mathcal{M}_{np}Q^{pm}B_{m}^{\mu\nu}\,+\, 
\epsilon_{qmr}Q^{mn}\left(  
                           \partial_{\mu}\mathcal{M}\cdot \mathcal{M}^{-1}
                     \right)_{n}{}^{q}\, A^{2\,r\,\mu} 
\right\}\; ,\\
\end{array}
\end{equation} 

%and upon using
%\begin{eqnarray}
%\left(
%  d\hat{k}_{(n)}
%\right)_{ai} &=& {e_{i}}^{m}\partial_{a}G_{nm} \\
%%
%\left(
%  d\hat{k}_{(n)}
%\right)_{ab} &=& G_{nm} \left(
%                            dA^{1\ m}
%                        \right)_{ab} \\
%%
%i_{\hat{k}_{(m)}}\hat{C}_{ab} &=& B_{m\ ab} \\
%%
%i_{\hat{k}_{(m)}}\hat{C}_{ai} &=& {e_{i}}^{p}\epsilon_{pmn}C_{(1)}^{n} \\
%%
%i_{\hat{k}_{(m)}}\hat{C}_{ij} &=& a {e_{i}}^{n} {e_{j}}^{p} \epsilon_{mnp} 
%%
%\end{eqnarray}

and the contorsion

\begin{equation}
 \label{eq:KKwadraat}
\begin{array}{rcl}
-\sqrt{|\hat{\hat{g}}|}\, 
\hat{\hat{K}}_{\hat{\hat{\mu}}\hat{\hat{\nu}}\hat{\hat{\rho}}}
\hat{\hat{K}}{}^{\hat{\hat{\mu}}\hat{\hat{\nu}}\hat{\hat{\rho}}}
& = & 
\sqrt{|g_{E}|}\, \left\{
  -\textstyle{\frac{1}{4}}e^{-\varphi} 
 B_{m\, \mu\nu}(Q\mathcal{M}Q)^{mn} B_{n}{}^{\mu\nu} 
\right. \\
& & \\
& & 
\hspace{-1cm}
+\textstyle{\frac{1}{2}}
     \left[
       (Q\mathcal{M}Q)^{mv}\mathcal{M}^{px}\epsilon_{pms}\epsilon_{xuw}
       + Q^{pv}Q^{mx}\epsilon_{pms}\epsilon_{xvw}
     \right] A^{2\, s}{}_{\mu}A^{2\,w\, \mu} \\
& & \\
& & 
\hspace{-2cm}
\left.
       -\textstyle{\frac{1}{4}}e^{\varphi}\ a^2
         \left[
            \left( Q\mathcal{M}Q\right)^{ms}\mathcal{M}^{ru}\mathcal{M}^{pv}
               \epsilon_{rpm}\epsilon_{uvs}
            -2Q^{mn}Q^{sp}\mathcal{M}^{ru}\epsilon_{mpr}\epsilon_{nsu}
         \right]
    \right\} \; .\\
\end{array}
\end{equation}

It can be seen that the terms in Eq. (\ref{eq:yoquese}) and the first
two terms in Eq. (\ref{eq:KKwadraat}) combine in just the right way
with the $F^{1\ n}$ and the $\partial\mathcal{M}\ \mathcal{M}^{-1}$
terms in Eq. (\ref{eq:curvature}), as to promote them to their
gauge-covariant equivalents in Eq. (\ref{eq:massivefieldstrengths}).

Note that one also finds a potential for $a$, which
however can be rewritten as
%one can make use of
%\begin{equation}
%\begin{array}{rclrcl}
%\mathcal{M}^{mr}\mathcal{M}^{ns}\mathcal{M}^{qt}\epsilon_{mnp} 
%    &=& \epsilon^{rst}\, ,\hspace{1cm} &
%\epsilon_{rmp}\epsilon^{rmp} &=& 3! \; , \\
%& & & & & \\
%\epsilon_{rmp}\epsilon^{rmq} &=& 2{\delta_{p}}^{q}\, ,&
%\epsilon_{rmp}\epsilon^{rnq} &=& 
%{\delta_{m}}^{n}{\delta_{p}}^{q} -{\delta_{m}}^{q}{\delta_{p}}^{n}\; ,\\
%\end{array}
%\end{equation} 
%to bring the potential term for $a$ to the form
%
\begin{equation}
+\sqrt{|g_{E}|}\, \textstyle{\frac{1}{2}} e^{\varphi}a^{2}
\left\{\left[{\rm Tr}\, \left( Q\mathcal{M}\right)\right]^{2}
-2{\rm Tr}\, \left( Q\mathcal{M}Q\mathcal{M} \right)
\right\} \; ,
\label{eq:Potential2}
\end{equation}

\noindent so that it can be combined with the result in 
Eq.~(\ref{eq:Potential1}) to complete the potential for the scalars in
the $d=8$ theory as

\begin{equation}
{\cal V} =
-{\textstyle\frac{1}{2}}
\frac{|\tau |^{2}}{\Im{\rm m} (\tau )}
\left\{\left[{\rm Tr}\, \left( Q\mathcal{M}\right)\right]^{2}
-2{\rm Tr}\, \left( Q\mathcal{M}Q\mathcal{M} \right)
\right\} \; .
\end{equation}

The complete $d=8$ massive action can be written as

\begin{equation}
\label{eq:d8massiveaction}
  \begin{array}{rcl}
S & = & {\displaystyle\int} d^{8}x \sqrt{|g_{E}|}\,
\left\{ 
R_{E} 
+\frac{1}{4}{\rm Tr}\left(\mathcal{D} {\cal M}{\cal M}^{-1}\right)^{2}
+\frac{1}{4}{\rm Tr}\left(\partial {\cal W}{\cal W}^{-1}\right)^{2}
 \right.\\
& & \\
& & 
\hspace{2cm}
-\frac{1}{4}F^{i\, m}{\cal M}_{mn}{\cal W}_{ij} F^{j\, n} 
+\frac{1}{2\cdot 3!} H_{m}{\cal M}^{mn} H_{n}
-\frac{1}{2\cdot 4!} e^{-\varphi} G^{2} 
-{\cal V} \\
& & \\
& & 
\hspace{1cm}
-\frac{1}{6^{3}\cdot 2^{4}}
{\textstyle\frac{1}{\sqrt{|g_{E}|}}}\, \epsilon
\left[GGa -8G H_{m} A^{2\, m} +12 G G_{(2)}^{m}B_{m}
-16 H_{m} G_{(2)}^{m} C
\right.\\
&& \\
& & 
\hspace{2cm}
-8G\partial a C
-8\epsilon^{mnp} H_{m}H_{n}B_{p} 
+2H_{m}Q^{mn}B_{n}\left(
                     Ca + 6 B_{p}A^{2\ p}
                  \right)
\\
&& \\
&&
\hspace{2cm}
-3 B_{m}Q^{mn}B_{n}\left(
                       Ga + 2H_{m}A^{2\ m} 
                     + 3 B_{m}G_{(2)}^{m}+2C\partial a
                   \right)
 \\
&& \\
&& 
\hspace{2cm}
+4C{f_{pq}}^{m}A^{2\ p}A^{2\ q}H_{m}
-12 C {f_{pq}}^{m} A^{2\ p}G_{(2)}^{q}B_{m}\\
& & \\
& & 
\hspace{2cm}
\left.\left.
+\textstyle{\frac{3^{2}\cdot 11}{2}}\left(
                                      B_{n}Q^{nm}B_{m}
                                    \right)^{2}a
-\textstyle{\frac{2^{2}\cdot 3^{3}\cdot 11}{5}}B_{n}Q^{nm}B_{m}
       {f_{pq}}^{r} A^{2\ p}A^{2\ q}B_{r}
\right]\right\} \, ,\\
\end{array}
\end{equation}

\noindent where we have introduced the abbreviation 

\begin{equation}
G_{(2)}^{m} \;=\; F^{2\ m}\,+\, a F^{1\ m} \; .
\end{equation}

This is evidently a set of $SO(3)$-gauged theories (a different gauged
theory for each choice of non-singular mass matrix $Q^{mn}$; the
$SO(3)$ gauge symmetry is lost for singular choices of mass matrix),
the vectors $A^{2\,m}$ associated to 11-dimensional membranes wrapped
in 2 cycles of the internal $T^{3}$ playing the role of $SO(3)$ gauge
fields.  At the same time these are theories with massive fields: we
find explicit mass terms for the 3 $B_{m}$s of the form
$B_{m}Q^{mn}B_{n}$.  Some of these terms are implicit in squares of
the $A^{1\,m}$ vector field strengths. These vectors can be completely
gauged away (``eaten'' by the 2-forms) and play the role of
St\"uckelberg fields. In other words: in the physical spectrum of the
theory there are no quantum excitations associated to those vector
fields and the quantum excitations associated to the 2-forms are
massive.

We can wonder how this can happen in supergravity theories since all
the fields in the supergravity multiplet should have the same mass
(zero). The reason why is that supersymmetry is partially and
spontaneously broken: massive supergravity theories as formally
invariant under a certain modification of the full supersymmetry
transformations of the massless theory. However, the vacuum of these
theories breaks part of the symmetry and in that vacuum the
supergravity multiplet becomes reducible into a massless supergravity
multiplet and massive matter multiplets. In this case, the theory is
written in a form in which all the gauge symmetries of the massless
theory are formally present, but in the vacuum those symmetries
responsible for the masslessness of the 2-forms are spontaneously
broken.

This set of theories transforms into itself under global
$Sl(3,\mathbb{Z})$ transformations and therefore they are an
$Sl(3,\mathbb{Z})$ multiplet of theories.

At this point it is important to compare this set of theories with the
$d=8$ $SU(2)$ gauged supergravity of Salam and Sezgin \cite{kn:SaSe}.
To make the comparison easier we can first use a mass matrix
proportional to the identity, the proportionality constant being the
coupling constant $g$. We can immediately see that the kinetic terms
in their action are identical to those in our action
Eq.~(\ref{eq:d8massiveaction}) up to the redefinition $\varphi_{\rm
  SS} =-\varphi_{\rm ours}$ but with the roles of the vector fields
$A^{1\,m}$ and $A^{2\,m}$ reversed. The second difference we see is in
the potential, which in their case (but in our notation) we can write
in the form

\begin{equation}
  \begin{array}{rcl}
{\cal V}_{\rm SS} & = & 
{\displaystyle\frac{1}{\Im{\rm m}(\tau)}}{\cal V}({\cal M})\, ,\\
& & \\
{\cal V}_{\rm ours} & = & 
{\displaystyle\frac{|\tau|^{2}}{\Im{\rm m}(\tau)}}{\cal V}({\cal M})\, ,\\
& & \\
{\cal V}({\cal M}) & = & 
-{\textstyle\frac{1}{2}}g^{2}
\left\{\left({\rm Tr}\,\mathcal{M}\right)^{2}
-2{\rm Tr}\, \left( \mathcal{M}^{2} \right)
\right\} \; .\\
\end{array}
\end{equation}

These two potentials and the vectors $A^{1\,m}$ and $A^{2\,m}$ are
related by the $Sl(2,\mathbb{Z})$ transformation 

\begin{equation}
 \Lambda =
\left(
  \begin{array}{cc}
0 & \,\,\,\,1\\
& \\
-1 & \,\,\,\,0\\
  \end{array}
\right)\, , 
\end{equation}

\noindent according to the rules 
Eqs.~(\ref{eq:sl2rules1},\ref{eq:sl2rules2}).

%%%%%%%%%%%%%%%%%%%%%%%%%%%%%%%%%%%%%%%%%%%%%%%%%%%%%%%%%%%%%%%%%%%%%%%

\section{Vacua}
\label{sec-vacua}

In this Section we want to study the simplest vacuum of the theory, the
one which presumably preserves more supersymmetry. A more detailed
study of the vacua of this theory will be presented elsewhere.

The scalar potential of this theory is $Sl(3,\mathbb{R})$ symmetric
but completely breaks the $Sl(2,\mathbb{R})$ global invariance of he
massless theory. The factor ${\cal V}({\cal M})$ has the typical form
of a potential for the $Sl(3,\mathbb{R})/SO(2)$ fields and acts as a
sort of Romans-like mass term for $a$. It is reasonable to expect that
the simplest vacua will be those that minimize the factor ${\cal
  V}({\cal M})$ and thus we will assume that ${\cal V}({\cal M})$ has
a minimum for some constant values ${\cal M}={\cal M}_{0}$, the value
of ${\cal V}({\cal M})$ at this minimum being denoted by ${\cal V}_{0}$.
Further we will take $a=0$ (the invariance under constant shifts of
$a$ is broken and we cannot simply take any constant value as in the
massless case).  We are left with non-trivial metric and $\varphi$
field with potential ${\cal V}_{0}e^{-\varphi}$. The only
equations of motion that remain to be solved are

\begin{equation}
\begin{array}{rcl}
R_{\mu\nu} +\frac{1}{2}\partial_{\mu}\varphi\partial_{\nu}\varphi
+\frac{1}{12}{\cal V}_{0}e^{-\varphi}g_{\mu\nu} & = & 0\, ,\\
& & \\
\nabla^{2}\varphi+\frac{1}{2}{\cal V}_{0}e^{-\varphi} & = & 0\, .\\
\end{array}
\end{equation}

It is clear that, unless ${\cal V}_{0}=0$ Minkowski spacetime is not
going to be a solution. In fact these equations are almost identical
to those appearing in Romans' 10-dimensional theory. This is not
surprising since our theory contains the dimensional reduction of
Romans'. Thus, we can look for a solution similar to the
10-dimensional D8-brane, i.e.~a domain-wall-type solution.  Choosing
the coordinate system in which the solution takes a simpler form, we
find

\begin{equation}
  \begin{array}{rcl}
ds^{2} & = & \left(z/\ell \right)^{2/3} g_{ij}dy^{i}dy^{j}-dz^{2}\, ,\\
& & \\
e^{\varphi} & = & {\cal V}_{0}  z^{2}\, ,\\
\end{array}
\end{equation}

\noindent where $i,j=0,\ldots,6$ and $\ell$ is an integration constant 
with dimensions of length and $g_{ij}$ is any Ricci-flat metric
depending on the domain-wall worldvolume coordinates $y^{i}$.

It should be clear that this solution is not the dimensional reduction
of the D8-brane since the internal metric corresponding to this
solution is isotropic whereas the internal metric of the D8-brane reduced
to 8 dimensions is not.

It would be interesting to find the unbroken supersymmetries of this
solution but the supersymmetry transformation rules of this theory are
not available yet.

More complicated vacua (in particular the reduced D8-brane) can be
found using a given parametrization of the $Sl(3,\mathbb{R})/SO(3)$
coset representative and will be presented elsewhere.

%%%%%%%%%%%%%%%%%%%%%%%%%%%%%%%%%%%%%%%%%%%%%%%%%%%%%%%%%%%%%%%%%%%%%%
\section{Conclusions}
\label{sec-conclusions}
%%%%%%%%%%%%%%%%%%%%%%%%%%%%%%%%%%%%%%%%%%%%%%%%%%%%%%%%%%%%%%%%%%%%%%

Following a recent observation made by Cowdall \cite{kn:cow} that the
9-dimensional massive supergravity given in \cite{kn:MO} contained the
$d=9$ $N=2$ gauged supergravity, we have investigated the relation
between massive and gauged supergravities in eight dimensions.
Starting from a proposed eleven dimensional massive supergravity
theory \cite{kn:BLO,kn:MO}, we derived, by normal dimensional
reduction, an eight dimensional massive theory. The resulting massive
theory can be interpreted as a gauged symmetry, since it is invariant
under local $SO(3)$, which is the R-symmetry of the corresponding
supertranslation algebra, transformations. Comparing this theory with
the known $N=2$ $d=8$ gauged supergravity theory found by Sezgin and
Salam \cite{kn:SaSe}, we find striking resemblance 8-dimensional
S~duality relates the axidilaton $\tau$ and the gauge fields of the
two theories. The theory found in this paper is then interpreted as
the bosonic sector of the gauged supergravity theory one would have
found if one had gauged the 3 vector fields associated to wrapped
membranes instead of the 3 KK vector fields (Salam and Sezgin's
choice).  Even though, in order to prove completely that we have
obtained the bosonic sector of a gauged supergravity theory, we need
to include the fermions and find the supersymmetry rules. It is very
reasonable to expect that both things can be done.

What are the implications of this result for the massive
11-dimensional supergravity theory of Refs.~\cite{kn:BLO,kn:MO}?  We
think we have shown that this theory is, at the very least, an
effective tool from which to obtain lower-dimensional massive
supergravities that cannot be obtained either by standard or by
Scherk-Schwarz generalized dimensional reduction. These theories seem
to have in common gauge vector fields associated to the 11-dimensional
3-form but their more important feature is the they are related by
some sort of lower-dimensional S~duality transformations to those
theories that can be obtained by more standard methods.

A possible interpretation of these results can be proposed based on
experience in the construction of new solutions using the mechanism
{\it reduction-S~dualization-oxidation} in the cases in which the
reduced theory has an S~duality symmetry that the original higher
dimensional theory lacks. The best-known example of the use of this
mechanism is the construction of the KK-monopole \cite{kn:So,kn:GrPe}
starting from the plane-wave solution in 5-dimensional gravity: the
plane-wave solution is reduced in the direction in which it propagates
and then S~dualized using the electric-magnetic duality symmetry
present in $d=4$ (but completely absent in the 5-dimensional theory!).
Finally, it is oxidized back to $d=5$ only to find that the new
solution generated in this way (the KK monopole) does not want to be
decompactified and can only live in a 5-dimensional space with a
compact dimension. It is possible to construct more solutions using
this mechanism and in general one finds that they have a number of
dimensions that cannot be decompactified \cite{kn:L-TO2}.

In Refs.~\cite{kn:BLO,kn:MO} it was argued that the massive
11-dimensional supergravity proposed in them could be interpreted as
the 11-dimensional supergravity one gets if one places KK9-branes
(referred to as M9-branes in Ref.~\cite{kn:BvdS}) in the vacuum. The
proposed KK9-branes are objects which necessarily live in
11-dimensional spacetime with 1 compact dimension.  On the other hand,
being $(d-2)$-branes, they are associated to mass parameters in the
action that can be interpreted as field strengths. As soon as these
mass parameters are non-vanishing in the action, one can say that the
vacuum contains the corresponding $(d-2)$-branes. This is different
from lower-dimensional branes: the presence of the field strengths of
the potentials to which they couple does not imply that they have to
have non-trivial values and there are vacuum solutions (Minkowski
spacetime, say) in which these field strengths vanish. Thus, if the
11-dimensional vacuum contains KK9-branes, then we expect mass
parameters in the action as well as explicit Killing vectors, since
these objects break 11-dimensional Poincar\'e invariance forcing one
dimension to be compact.

It is clear from all this that the KK9-branes (or, perhaps, objects
with more isometric dimensions) are produced when the {\it
  reduction-S~dualization-oxidation} mechanism is used on
11-dimensional supergravity. The resulting theory, which we have called
massive 11-dimensional supergravity cannot be decompactified.

Clearly more work is needed to confirm the interpretation of the
massive 11-dimensional supergravity put forward here. In particular
one would like to prove that the theory (or the family of theories one
gets with $N$ Killing vectors and arbitrary mass matrices) can be
supersymmetrized directly in eleven dimensions even if under the
condition of the existence of $N$ isometries.  It would also be very
interesting to see if these further straightforward generalizations to
$N$ Killing vectors lead to new gauged supergravities in lower than 8
dimensions.  Especially the case with 6 Killing vectors should be
interesting since then we could compare it to type IIB compactified on
$S^{5}$.  Work in these directions is in progress.

%%%%%%%%%%%%%%%%%%%%%%%%%%%%%%%%%%%%%%%%%%%%%%%%%%%%%%%%%%%%%%%%%%%%%%
\section*{Acknowledgments}
%%%%%%%%%%%%%%%%%%%%%%%%%%%%%%%%%%%%%%%%%%%%%%%%%%%%%%%%%%%%%%%%%%%%%%

The work of N.A.-A.~and T.O.~is supported by the European Union TMR
program FMRX-CT96-0012 {\sl Integrability, Non-perturbative Effects,
  and Symmetry in Quantum Field Theory} and by the Spanish grant
AEN96-1655.  The work of P.M. was partially supported by the
F.W.O.-Vlaanderen and the E.U.~RTN program HPRN-CT-2000-00131.

%%%%%%%%%%%%%%%%%%%%%%%%%%%%%%%%%%%%%%%%%%%%%%%%%%%%%%%%%%%%%%%%%%%%%%
\appendix
%%%%%%%%%%%%%%%%%%%%%%%%%%%%%%%%%%%%%%%%%%%%%%%%%%%%%%%%%%%%%%%%%%%%%%
\section{$SO(3)$ in a Non-Standard Basis}
%%%%%%%%%%%%%%%%%%%%%%%%%%%%%%%%%%%%%%%%%%%%%%%%%%%%%%%%%%%%%%%%%%%%%%
\label{sec-so3}

The standard basis for the Lie algebra of $SO(3)$ with anti-Hermitean
generators $\{T_{m}\}$,  $m=1,2,3$ is such that

\begin{equation}
[T_{m},T_{n}] =\epsilon_{mnp}T_{p}\, ,  
\end{equation}

\noindent so the structure constants are given by $f_{mn}{}^{p}
=\epsilon_{mnp}$. The Killing metric for the adjoint representation
${\rm Adj}\left(T_{m}\right)^{p}{}_{n}=\epsilon_{mnp}$ is just

\begin{equation}
K_{mn}={\rm Tr}\left[{\rm Adj}\left(T_{m}\right)
{\rm Adj}\left(T_{n}\right)  \right] = -2\delta_{mn}\, .
\end{equation}

Let us now perform a change of basis and compute the new structure constants:

\begin{equation}
T^{\prime}_{m}=R_{m}{}^{n}T_{n}\, ,
\hspace{1cm}
[T^{\prime}_{m},T^{\prime}_{n}] =R_{m}{}^{r}R_{n}^{s}\epsilon_{rsp}T_{p}=
{\rm det}R\, \epsilon_{mnq}\left(R^{-1}\right)_{p}{}^{q}
\left(R^{-1}\right)_{p}{}^{r} T^{\prime}_{r}\, ,
\end{equation}

\noindent where we have used that $R$ is an arbitrary non-singular matrix. 
Thus

\begin{equation}
f_{mn}{}^{p}=\epsilon_{mnq} Q^{qp}\, ,
\hspace{1cm}
Q^{qp}= Q^{pq}  = {\rm det}R \left(R^{-1}\right)_{s}{}^{p}
\left(R^{-1}\right)_{s}{}^{q}\, .
\end{equation}

The Killing metric in the new basis is simply related to the symmetric
matrix $Q^{mn}$ by

\begin{equation}
K_{mn} = {\rm det}Q\, (Q^{-1})_{mn}\, .
\end{equation}

Fields in the fundamental (vector or adjoint) representation of
$SO(3)$ behave under infinitesimal transformations with parameter

\begin{equation}
\sigma^{m}{}_{n}\equiv\sigma^{p}\left(T_{p}\right)^{m}{}_{n}=
\sigma^{p}f_{pn}{}^{m}=\sigma^{p}\epsilon_{pnq}Q^{qm}\, ,  
\end{equation}

\noindent either contravariantly or covariantly

\begin{equation}
\left\{
\begin{array}{rcl}
\delta \psi^{m} & = & \sigma^{m}{}_{n}\psi^{n}\, ,\\
& & \\
\delta \xi_{m} & = & -\xi_{n}\sigma^{n}{}_{m}\, ,\\
\end{array}
\right.
\end{equation}

\noindent and their covariant derivatives are defined by\footnote{We 
  have absorbed the gauge coupling constant into the redefinition of
  the Lie algebra generators, {\em i.e.}~into the matrix $Q^{mn}$.}

\begin{equation}
\left\{
\begin{array}{rcl}
{\cal D}_{\mu}\psi^{m} & = & 
\partial_{\mu}\psi^{m}-f_{np}{}^{m}A^{n}{}_{\mu}\psi^{p}\, ,\\
& & \\
{\cal D}_{\mu}\xi_{m} & = & 
\partial_{\mu}\xi_{m}+f_{nm}{}^{p}A^{n}{}_{\mu}\xi_{p}\, ,\\
\end{array}
\right.
\end{equation}

\noindent where the gauge field transforms according to

\begin{equation}
\delta A^{m}{}_{\mu}= {\cal D}_{\mu}\sigma^{m}= 
\partial_{\mu}\sigma^{m}-f_{np}{}^{m}A^{n}{}_{\mu}\sigma^{p}\, .
\end{equation}

The gauge field strength is given by

\begin{equation}
F^{m}{}_{\mu\nu} = 2\partial_{[\mu}A^{m}{}_{\nu]} 
-f_{np}{}^{p}A^{n}{}_{\mu}A^{p}{}_{\nu}=
 2\partial_{[\mu}A^{m}{}_{\nu]} 
-Q^{mq}\epsilon_{npq}A^{n}{}_{\mu}A^{p}{}_{\nu}\, .   
\end{equation}

%%%%%%%%%%%%%%%%%%%%%%%%%%%%%%%%%%%%%%%%%%%%%%%%%%%%%%%%%%%%%%%%%%%%%%

\end{document}